\documentclass[12pt]{article}
\usepackage{amsmath}
\usepackage{amssymb}
\usepackage{amsthm}
\usepackage{graphicx}
\usepackage{mathrsfs}
\usepackage{cite}
\usepackage[margin=1in]{geometry}
\usepackage{url}
\usepackage{hyperref}
\title{Spin and the Thermal Equilibrium \\Distribution of Wave Functions}
\author{Viraj Pandya\footnote{Mathematics \& Economics, Class of 2013, Rutgers University, Serin Physics \& Astronomy Building, 126 Frelinghuysen Road, Piscataway, NJ 08854-8019, USA. E-mail: viraj.pandya@gmail.com}, Roderich Tumulka\footnote{To whom correspondence should be addressed}\ \footnote{Department of Mathematics, Rutgers University, Hill Center, 110 Frelinghuysen Road, Piscataway, NJ 08854-8019, USA. E-mail: tumulka@math.rutgers.edu}}
\date{August 30, 2013}

\theoremstyle{plain}\newtheorem{prop}{Proposition}
\newcommand{\be}{\begin{equation}}
\newcommand{\ee}{\end{equation}}
\newcommand{\Hilbert}{\mathscr{H}}
\newcommand{\sphere}{\mathbb{S}}

\newcommand{\EEE}{\mathbb{E}}
\newcommand{\PPP}{\mathbb{P}}
\DeclareMathOperator{\tr}{tr}
\DeclareMathOperator{\Var}{Var}
\renewcommand{\Re}{\mathrm{Re}}
\renewcommand{\Im}{\mathrm{Im}}
\newcommand{\red}{\mathrm{red}}
\newcommand{\cond}{\mathrm{cond}}
\newcommand{\scp}[2]{\langle #1|#2\rangle}

\begin{document}
\maketitle

\abstract{
Consider a quantum system $S$ weakly interacting with a very large but finite system $B$ called the heat bath, and suppose that the composite $S\cup B$ is in a pure state $\Psi$ with participating energies between $E$ and $E+\delta$ with small $\delta$. Then, it is known that for most $\Psi$ the reduced density matrix of $S$ is (approximately) equal to the canonical density matrix. That is, the reduced density matrix is universal in the sense that it depends only on $S$'s Hamiltonian and the temperature but not on $B$'s Hamiltonian, on the interaction Hamiltonian, or on the details of $\Psi$. It has also been pointed out that $S$ can also be attributed a random wave function $\psi$ whose probability distribution is universal in the same sense. This distribution is known as the ``Scrooge measure'' or ``Gaussian adjusted projected (GAP) measure''; we regard it as the thermal equilibrium distribution of wave functions. The relevant concept of the wave function of a subsystem is known as the ``conditional wave function.'' In this paper, we develop analogous considerations for particles with spin. One can either use some kind of conditional wave function or, more naturally, the ``conditional density matrix,'' which is in general different from the reduced density matrix. We ask what the thermal equilibrium distribution of the conditional density matrix is, and find the answer that for most $\Psi$ the conditional density matrix is (approximately) deterministic, in fact (approximately) equal to the canonical density matrix. 

\medskip

}

\section{Introduction and Overview}
In this paper, we review the thermal equilibrium distribution of wave functions \cite{gap2006,smooth2005,Rei08,GLMTZ11}, known as the ``Scrooge measure'' or the ``$GAP$ measure,'' and discuss how the derivation of this distribution changes when spin is taken into account. In fact, it is the distribution of the \emph{conditional wave function} $\psi^\cond_S$ of a system $S$ that converges to $GAP$ in thermal equilibrium, and the relevant difference between particles with and without spin concerns $\psi^\cond_S$, as follows. The definition of $\psi^\cond_S$ for a system $S$ entangled with another system $B$ is based on the wave function $\Psi\in\Hilbert=\Hilbert_S\otimes \Hilbert_B$ of the composite $S\cup B$ and on a choice of generalized orthonormal basis (GONB)\footnote{By a GONB we mean that which is provided by a unitary isomorphism $\Hilbert_B\to L^2(\mathscr{Y})$ for some measure space $\mathscr{Y}$ containing the $y$; this includes the possibility of a continuous basis such as the position basis.} $\{|y\rangle\}$ in $\Hilbert_B$; namely, one picks a $|y\rangle$ at random with the appropriate marginal of the $|\Psi|^2$ distribution, say $|Y\rangle$, forms the partial inner product $\scp{Y}{\Psi}$, and normalizes the resulting vector in $\Hilbert_S$ to obtain $\psi^\cond_S$. The usual choice of $\{|y\rangle\}$ is the position basis, and this is fine for spinless particles but in the presence of spin it is not a basis. In the latter case, one possibility is to take for $\{|y\rangle\}$ a product basis of the position basis and an orthonormal basis in spin space, such as, for each particle, $\bigl|\uparrow\bigr\rangle$ and $\bigl|\downarrow\bigr\rangle$; a drawback of this choice is that it prefers one direction in space, the $z$ direction. Another, perhaps more natural, possibility is to use the \emph{conditional density matrix} $\rho^\cond_S$ instead of $\psi^\cond_S$, and this possibility will be explored here. This notion refers to the situation in which $\Hilbert_B=\Hilbert_y\otimes\Hilbert_s$ (so that, in particular, $\Hilbert_y$ can be the spatial degrees of freedom and $\Hilbert_s$ the spin degrees of freedom of system $B$) and is based on a GONB $\{|y\rangle\}$ of $\Hilbert_y$ and a wave function $\Psi\in\Hilbert_S\otimes\Hilbert_y\otimes\Hilbert_s$; one picks $|Y\rangle$ at random with the appropriate marginal of the $|\Psi|^2$ distribution,
forms $\psi^\cond_{S\cup s}\propto\scp{Y}{\Psi}\in\Hilbert_S\otimes\Hilbert_s$, and then sets
\be\label{rhoconddef}
\rho^\cond_S = \tr_s \Bigl|\psi^\cond_{S\cup s}\Bigr\rangle \Bigl\langle \psi^\cond_{S\cup s}\Bigr|\,,
\ee
a density matrix on $\Hilbert_S$. Note that $Y$ is not averaged over; instead, $\rho^\cond_S$ depends on $Y$ and is therefore random.

The difference between the two options, of either introducing a basis in spin space or using the conditional density matrix, becomes particularly
salient in Bohmian mechanics \cite{DGZ92,dm2005}. In that framework, the conditional wave function is obtained by inserting the \emph{actual}
$y$-configuration into $\Psi$; this procedure cannot be repeated with spin because the Bohmian particles, although they have actual positions, do not
have actual spin values. Thus, if system $B$ involves spin, the quantity provided by Bohmian mechanics as the state of $S$ is the conditional
density matrix (with the actual $y$-configuration inserted and $s$ traced out) rather than the conditional wave function \cite{dm2005}. Outside the
Bohmian framework, the two options are perhaps on the same footing, and the choice is a matter of taste.

Our new contribution here is an investigation of the distribution of $\rho^\cond_S$ for $S$ in thermal equilibrium with a (large but finite) heat bath $B$ consisting of particles with spin. We begin with an overview of canonical typicality and the thermal equilibrium distribution of wave functions.

\subsection{Canonical Typicality}

We consider two quantum systems, $S$ (``the system'') and $B$ (``the heat bath''), such that $B$ is very large (i.e., is much bigger than $S$ and has at least, say, $10^{20}$ particles); suppose that $S$ is entangled with $B$, that the composite system $S\cup B$ is in a pure state $\Psi\in\Hilbert_{S\cup B}=\Hilbert_S \otimes \Hilbert_B$, and that $S\cup B$ is isolated, so that $\Psi$ evolves according to the Schr\"odinger equation with Hamiltonian
\be\label{Hdef}
H_{S\cup B}=H_S\otimes I_B + I_S\otimes H_B + H_\mathrm{interaction}\,,
\ee
where $I$ denotes the identity operator. We assume that $S\cup B$ is confined to a finite volume, so that $H_{S\cup B}$ has pure point spectrum. Let $[E,E+\delta]$ be an energy interval for $S\cup B$ that is small on macroscopic scales but large enough to contain a great (but finite) number of eigenvalues of $H_{S\cup B}$,
and let $\Hilbert_{[E,E+\delta]}$ be the corresponding spectral subspace, i.e., the subspace of $\Hilbert_{S\cup B}$ spanned by the eigenvectors of $H_{S\cup B}$ with eigenvalues in $[E,E+\delta]$. Let the \emph{micro-canonical density matrix} be 
\be
\rho_{[E,E+\delta]} = (\dim\Hilbert_{[E,E+\delta]})^{-1} P_{\Hilbert_{[E,E+\delta]}} \; ,
\ee
where $P_{W}$ denotes the projection to the subspace $W$, and let the \emph{micro-canonical distribution} $u_{[E,E+\delta]}$ be the uniform probability distribution on $\sphere(\Hilbert_{[E,E+\delta]})$, where $\sphere$ denotes the unit sphere,
\be\label{spheredef}
\sphere(\Hilbert) = \bigl\{ \psi\in\Hilbert: \|\psi\|=1 \bigr\}\,.
\ee

It has long been known (e.g., \cite{tolman}) that, if the interaction term in \eqref{Hdef} is negligibly small, then, in the thermodynamic limit (i.e., as the size or number of components $N$ of $B$ goes to infinity and $E/N \rightarrow e<\infty$), the partial trace $\tr_B\rho_{[E,E+\delta]}$ approaches the canonical density matrix,
\be
\rho_{\beta}= \frac{1}{Z}\exp(-\beta H_S) \,,
\ee
where $Z=\tr \exp(-\beta H_S)$ is the normalization factor. A stronger statement is, in fact, true:

\begin{prop}\label{prop:cantyp}
(Canonical typicality \cite{GMM04,cantyp2006, PSW05,PSW06}) Suppose that the interaction between $S$ and $B$ is negligible, that the dimensions of $\Hilbert_B$ and $\Hilbert_{[E,E+\delta]}$ are sufficiently large, and that $H_B$ has a reasonable distribution of eigenvalues. Then, for most wave functions $\Psi\in\sphere(\Hilbert_{[E,E+\delta]})$, the reduced density matrix $\rho_S^{\red}=\tr_B |\Psi\rangle \langle\Psi|$ of $S$ is approximately canonical: $\rho^\red_S \approx \rho_\beta$ for suitable $\beta$.
\end{prop}

The statement that \emph{most} $\Psi\in\sphere(\Hilbert_{[E,E+\delta]})$ have a property $p$ means that the measure $u_{[E,E+\delta]}$ of the set of $\Psi$ with property $p$ is close to 1; how close, and the degree of closeness in ``$\approx$'', depend on $\dim\Hilbert_{[E,E+\delta]}$; see \cite{PSW05} for explicit error bounds. The condition that we vaguely call ``reasonable distribution of eigenvalues'' means that one can reasonably speak of a density of states (i.e., distribution density of eigenvalues) that is a differentiable function. This condition is needed already for concluding that $\tr_B \rho_{[E,E+\delta]}\approx \rho_\beta$.

We summarize Proposition~\ref{prop:cantyp} by saying that when $S\cup B$ is in thermal equilibrium then $\rho^\red_S$ is canonical. The connection between thermal equilibrium and the typical properties of $\Psi$ (i.e., the properties shared by \emph{most} $\Psi$ in the energy shell), a connection that could even be taken as the definition of thermal equilibrium, can be understood by noting that, for a typical property $p$, the time evolution (with nonzero interaction term) should sooner or later lead to a $\Psi_t$ which has property $p$, and in fact that $\Psi_t$ should have property $p$ for most times $t$ in the long run---which is the behavior characteristic of thermal equilibrium.

\subsection{Thermal Equilibrium Distribution of the Conditional Wave Function}

Our next proposition goes beyond Proposition~\ref{prop:cantyp} and involves the $GAP$ (Gaussian adjusted projected) measures \cite{JRW94,gap2006,Rei08}. We give the definition of $GAP$ measures in Section~\ref{sec:GAPdef} and note at this point only that for every Hilbert space $\Hilbert$ and every density matrix $\rho$ on $\Hilbert$ there is a measure $GAP(\rho)$; it is a probability distribution over $\sphere(\Hilbert)$.

The proposition about $GAP$ measures concerns the wave function of $S$, the precise notion of which is the conditional wave function $\psi_S^\cond$. This notion was first defined in \cite{DGZ92} for Bohmian mechanics as
\be
\psi_S^\cond(x) = \mathcal{N}^{-1} \Psi(x,Y)\,,
\ee
where both $\Psi$ and $\psi_S^\cond$ are expressed in the position representation (of, say, spinless particles), $\mathcal{N}$ is the normalizing factor,\footnote{This factor, $\mathcal{N}=\|\Psi(\cdot,Y)\|$, will fail to be well defined if $\Psi(\cdot,Y)$ fails to be square-integrable. However, the $Y$ for which this happens form a set of measure zero because $\int dy \, \|\Psi(\cdot,y)\|^2 = \int dy\int dx\, |\Psi(x,y)|^2<\infty$. The factor $\mathcal{N}$ could be zero, but since $Y$ has distribution density $\|\Psi(\cdot,y)\|^2$, also this case occurs with probability zero.} $x$ is the configuration variable for $S$, and $Y$ is the actual (Bohmian) configuration of $B$. We use here a more abstract definition \cite{GN99,gap2006} that is independent of the Bohmian framework and fits nicely for our purposes: Let $\{|y\rangle\}$ be an orthonormal basis (ONB) of $\Hilbert_B$.\footnote{If $\dim\Hilbert_B=\infty$, we can also admit a GONB. Since we assume here that $\dim\Hilbert_B<\infty$, every GONB is an ONB.} Let $|Y\rangle$ be a randomly chosen element of the ONB with probability distribution
\be
\PPP(Y=y) = \bigl\| \scp{y}{\Psi} \bigr\|^2\,,
\ee
where the inner product is a partial inner product and thus a vector in $\Hilbert_S$, and $\|\cdot\|$ denotes the norm in $\Hilbert_S$. Then
\be\label{psiconddef}
\psi_S^\cond = \mathcal{N}^{-1} \scp{Y}{\Psi}
\ee
with $\mathcal{N}= \bigl\| \scp{Y}{\Psi} \bigr\|$. Note that the conditional wave function is a \emph{random} vector in $\Hilbert_S$.

\begin{prop}\label{prop:GAP}
(Thermal equilibrium distribution of the conditional wave function \cite{gap2006,GLMTZ11}) Suppose that the interaction between $S$ and $B$ is negligible, that the dimensions of $\Hilbert_B$ and $\Hilbert_{[E,E+\delta]}$ are sufficiently large, and that $H_B$ has a reasonable distribution of eigenvalues. Then, for most wave functions $\Psi\in\sphere(\Hilbert_{[E,E+\delta]})$ and most ONBs $\{|y\rangle\}$ of $\Hilbert_B$, the distribution $\mu^\cond_S$ of the conditional wave function $\psi^\cond_S$ is approximately $GAP$: $\mu^\cond_S\approx GAP(\rho_\beta)$, with the same value of $\beta$ as in Proposition~\ref{prop:cantyp}.
\end{prop}

Talk about ``most'' ONBs refers to the uniform probability distribution over the set of all ONBs of $\Hilbert_B$, a distribution closely related to the Haar measure on the unitary group of $\Hilbert_B$. Also, the degrees of closeness in ``most'' and ``$\approx$''  again depend on $\dim \Hilbert_{[E,E+\delta]}$; see \cite{GLMTZ11} for explicit error bounds and a precise definition of ``$\approx$.''

\subsection{Spin and the Conditional Density Matrix}

We now assume that the particles belonging to $B$ (as well as those of $S$) have spin. The state vector $\Psi$ of $S\cup B$ can then be written as a wave function $\Psi_{r,s}(x,y)$, where $x$ is the configuration of $S$, $r$ is a cumulative index for the spin degrees of freedom of $S$, $y$ is the configuration of $B$, and $s$ is the cumulative index for the spin degrees of freedom of $B$. Put differently,
\be\label{Psirsxy}
\Psi \in \Hilbert_x \otimes \Hilbert_r \otimes \Hilbert_y \otimes \Hilbert_s \; ,
\ee
where $\Hilbert_x\otimes\Hilbert_r=\Hilbert_S$ and $\Hilbert_y\otimes\Hilbert_s=\Hilbert_B$. The conditional wave function in the framework of Bohmian mechanics is then proportional to $\Psi_{r,s}(x,Y)$, which has more spin indices than belong to $S$. 

For this reason, the conditional density matrix has been introduced \cite{dm2005}, 
\be\label{rhocondrx}
\rho^{\cond}_{S\:\:\:r,r'} (x,x') = \frac{1}{\mathcal{N}(Y)}\sum_{s} \Psi_{r,s}(x,Y)\, \Psi^*_{r',s}(x',Y) \; ,
\ee
where
\be
\mathcal{N}(Y)=\sum_{r,s}\int dx \, \bigl|\Psi_{r,s}(x,Y)\bigr|^2
\ee
is the normalizing factor; $\rho^\cond_S$ is a density matrix on $\Hilbert_S=\Hilbert_x\otimes\Hilbert_r$. The more abstract formulation analogous to our definition \eqref{psiconddef} of the conditional wave function and independent of the Bohmian framework has already been given around \eqref{rhoconddef} above. Equivalently,
\be\label{rhoconddef2}
\rho^\cond_S=\frac{\tr_s \scp{Y}{\Psi} \scp{\Psi}{Y}}{\bigl\| \scp{Y}{\Psi}\bigr\|^2}\,.
\ee

We make a few general remarks about the conditional density matrix. First, $\rho^\cond_S$ is random because $Y$ is; its distribution $\nu_S^\cond$ depends on $\Psi$. Second, the average of $\rho^\cond_S$ is the reduced density matrix,
\be
\EEE_Y \rho^\cond_S = \sum_y \tr_s \scp{y}{\Psi} \scp{\Psi}{y} = \tr_{y\cup s} |\Psi\rangle\langle\Psi| = \rho^\red_S\,.
\ee
Third, if we choose an ONB $\{|s\rangle\}$ of $\Hilbert_s$ and form the conditional wave function $\psi^\cond_S$ (as opposed to $\psi^\cond_{S\cup s}$!), then $\rho^\cond_S$ can be obtained from $|\psi^\cond_S\rangle\langle\psi^\cond_S|$ by averaging over $s$ but not over $y$ (i.e., by taking the conditional expectation, given $Y$).

\subsection{Thermal Equilibrium Distribution of the Conditional Density Matrix}

We will show that, like $\psi^\cond_S$, also $\rho^\cond_S$ has a universal distribution in thermal equilibrium (provided that the coupling between $S$ and $B$ is weak); that is, that $\nu_S^\cond$ is nearly independent of $\Psi$ (i.e., it is approximately the same distribution for most $\Psi$) in the energy shell, as well as independent of $H_B$. However, it may be surprising what this universal distribution is. For example, one might have expected the thermal distribution of the conditional density matrix to be an exponential variant of the Gaussian adjusted projected measure for $\psi^\cond_S$: perhaps, an exponential adjusted projected measure over all density matrices. Instead, it turns out that the universal distribution is a delta peak; that is, $\rho^\cond_S\approx \rho_\beta$ with probability near 1.

\begin{prop}\label{prop:condDM}
(Thermal distribution of the conditional density matrix) Suppose that the interaction between $S$ and $B=y\cup s$ is negligible, as well as that between $y$ and $s$, that the dimensions of $\Hilbert_y$, $\Hilbert_s$, and $\Hilbert_{[E,E+\delta]}$ are sufficiently large, and that $H_y$ and $H_s$ have reasonable distributions of eigenvalues. Then, for most wave functions $\Psi\in\sphere(\Hilbert_{[E,E+\delta]})$ and most ONBs $\{|y\rangle\}$ of $\Hilbert_y$, the distribution $\nu^\cond_S$ of the conditional density matrix $\rho_S^{cond}$ is narrowly peaked at the canonical density matrix: $\nu^\cond_S \approx \delta_{\rho_\beta}$ with the same value of $\beta$ as in Proposition~\ref{prop:cantyp}.
\end{prop}

Here, the notation $\delta_x$ means the probability distribution concentrated on the single point $x$. Again, the degrees of closeness in ``most'' and ``$\approx$'' depend on $\dim \Hilbert_{[E,E+\delta]}$; we describe an estimate of the narrowness of the peak in Section~\ref{sec:CLT}. 

Proposition~\ref{prop:condDM} is the main result of this paper. As already mentioned in the beginning, there are obvious alternatives to using the conditional density matrix; in fact, the concept of conditional wave function can be used in two different ways, both of which lead to $GAP$ distributions. One way is to use the conditional wave function $\psi^\cond_{S\cup s}\in\Hilbert_x\otimes \Hilbert_r \otimes \Hilbert_s$. Alternatively, a basis could be chosen in $\Hilbert_s$, and then the conditional wave function $\psi^\cond_S\in\Hilbert_x \otimes \Hilbert_r$ relative to that basis could be used. 

The physical meaning of Proposition~\ref{prop:condDM} becomes particularly clear in Bohmian mechanics. In that framework, if system $B$
involves spin, then the physical state of system $S$ is given by $\rho^\cond_S$ (together with the actual $x$-configuration), and
Proposition~\ref{prop:condDM} describes the probability distribution of the physical state of system $S$.

The remainder of this paper is organized as follows. In Section 2, we review the $GAP$ measure. In Section 3, we derive our main result, the thermal equilibrium distribution of the conditional density matrix.

\section{The $GAP(\rho)$ Measure}

In this section, we define and discuss the Gaussian adjusted projected ($GAP$) measures, called the Scrooge measures in \cite{JRW94}. We write $X\sim Y$ to indicate that the random variables $X$ and $Y$ have the same distribution, and $X\sim \mu$ to indicate that $X$ has distribution $\mu$.

\subsection{Definition of the $GAP$ Measure}\label{sec:GAPdef}

Let $\Hilbert$ be a Hilbert space and $\rho$ a density matrix on $\Hilbert$ (i.e., a positive operator with $\tr \rho=1$). We describe three equivalent definitions of $GAP(\rho)$. 

The first involves Gaussian measures and proceeds in three steps represented by the acronym $GAP$. Let $G(\rho)$ denote the Gaussian measure on $\Hilbert$ with covariance matrix $\rho$; it can be defined explicitly as follows. We call a complex random variable $X$ Gaussian with variance $\sigma^2$ iff $\Re\,X$ and $\Im\,X$ are independent real random variables, each with a Gaussian distribution with mean 0 and variance $\sigma^2/2$. Using the spectral decomposition
\be
\rho=\sum_j p_j \, |j\rangle \langle j| \,,
\ee
where $p_j$ are the eigenvalues of $\rho$ and $\{|j\rangle\}$ is an ONB of eigenvectors, and using a sequence $X_j$ of independent complex Gaussian random variables with variances $\sigma_j^2=p_j$, the random vector
\be
\Psi^{G(\rho)}=\sum_j X_j |j\rangle
\ee
has distribution $G(\rho)$. (If $\dim\Hilbert=\infty$, the series still converges in $\Hilbert$ because
\be\label{Gnorm2}
\EEE\sum_j |X_j|^2 = \sum_j\EEE|X_j|^2 = \sum_j p_j = \tr \rho = 1 \; ,
\ee
where $\EEE$ means expectation.)

The second step of the construction, the ``adjustment,'' consists of reweighting the measure by means of a density function $f:\Hilbert\to[0,\infty)$, namely $f(\psi)=\|\psi\|^2$. That is, the measure $GA(\rho)$ is defined by
\be\label{GAdef}
GA(\rho)(d\psi) = \|\psi\|^2 G(\rho)(d\psi)\, .
\ee
It is a probability measure by virtue of \eqref{Gnorm2}.

The third step is to project this measure to the unit sphere $\sphere(\Hilbert)$. That is, if $\Psi^{GA(\rho)}$ is a random vector with distribution $GA(\rho)$ then
\be
\Psi^{GAP(\rho)} = \frac{\Psi^{GA(\rho)}}{\|\Psi^{GA(\rho)}\|} \; .
\ee
has distribution $GAP(\rho)$. This completes the definition of the $GAP$ measures.

The following alternative definition \cite{JRW94} does not mention Gaussian measures; it applies when $n:=\dim\Hilbert<\infty$. Let $\Psi^u$ be uniformly distributed on $\sphere(\Hilbert)$, and let $D(\rho)$ denote the distribution of
\be
\Psi^{D(\rho)} = \sqrt{n\rho} \,\Psi^u\,.
\ee
$D(\rho)$ is a measure on $\Hilbert$ concentrated on the ellipsoid that is the image of the unit sphere under $\sqrt{n\rho}$. Then (as we will confirm in the next section)
\be\label{DAP=GAP}
DAP(\rho)=GAP(\rho)\,.
\ee
That is, to obtain $GAP(\rho)$ we apply the same adjust-and-project procedure as before to $D(\rho)$. This is possible because the function $f(\psi)=\|\psi\|^2$ has mean 1 under $D(\rho)$:
\be
\EEE\|\Psi^{D(\rho)}\|^2 = n\, \EEE \scp{\Psi^u}{\rho|\Psi^u} = n\EEE\tr\Bigl(\rho\, |\Psi^u\rangle\langle\Psi^u|\Bigr) = n \tr(\rho\,  n^{-1}I) = \tr \rho =1\,.
\ee

The following, third definition of $GAP(\rho)$ was suggested to us by an anonymous referee; like the previous definition, it applies when $n=\dim\Hilbert<\infty$. Let $\Hilbert_2$ be any Hilbert space of the same dimension $n$, and fix any vector $\Phi\in\sphere(\Hilbert\otimes\Hilbert_2)$ such that $\tr_2 |\Phi\rangle\langle\Phi|=\rho$. Select $\Psi_2\in\sphere(\Hilbert_2)$ at random with distribution 
\be
\mu_2(d\psi_2) = n\,  \bigl\|\scp{\psi_2}{\Phi}\bigr\|^2\, u_2(d\psi_2)\,,
\ee
where $\scp{\cdot}{\cdot}$ is a partial inner product, $\|\cdot\|$ is the norm in $\Hilbert$, and $u_2$ is the uniform probability distribution on $\sphere(\Hilbert_2)$; $\mu_2$ is normalized because
\be
\mu_2(\sphere(\Hilbert_2)) = n\int\limits_{\sphere(\Hilbert_2)} \!\!\! u(d\psi_2) \, \scp{\Phi}{\psi_2} \scp{\psi_2}{\Phi}
= n \,\scp{\Phi}{n^{-1}I_2|\Phi}=1\,.
\ee
Then $GAP(\rho)$ is the distribution of
\be
\Psi= \frac{\scp{\Psi_2}{\Phi}}{\bigl\| \scp{\Psi_2}{\Phi} \bigr\|}\,,
\ee
where $\scp{\cdot}{\cdot}$ is again a partial inner product. The equivalence of these definitions will become clear in the next section.

\subsection{Properties of the $GAP$ Measure}

With every probability distribution $\mu$ on $\sphere(\Hilbert)$ is associated the density matrix
\be\label{rhomudef}
\rho_{\mu} = \int\limits_{\sphere(\Hilbert)} \!\!\! \mu(d\psi) \, |\psi\rangle \langle\psi| \,.
\ee
The density matrix associated with $GAP(\rho)$ is $\rho$,
\be
\rho_{GAP(\rho)}=\rho\,.
\ee
To see this, note that \eqref{rhomudef} can be regarded as a special case of the covariance matrix whenever $\mu$ has mean 0, and that the covariance matrix can be defined also for probability measures $\mu$ on $\Hilbert$ (as opposed to $\sphere(\Hilbert)$) with mean 0 by
\be
C_\mu =\int\limits_{\Hilbert} \mu(d\psi) \, |\psi\rangle \langle\psi| \,.
\ee
Note further that the adjust-and-project procedure preserves the covariance matrix, $C_{\mu AP}=C_\mu$ \cite{gap2006}. Thus, $\rho_{GAP(\rho)}=C_{GAP(\rho)}=C_{G(\rho)}=\rho$.

It follows also that the second definition given above is equivalent to the first: In finite dimension $n$, $\Psi^{G(I/n)}= \Lambda \Psi^u$, where $\Lambda=\|\Psi^{G(I/n)}\|$ is a real-valued random variable independent of $\Psi^u$ with $\EEE\Lambda^2=1$. 
Note that $\Psi^{G(\rho)}\sim\sqrt{n\rho}\, \Psi^{G(I/n)}=\Lambda \sqrt{n\rho}\,\Psi^u$. The adjustment factor $f(\psi)$ can be written as 
$\Lambda^2 \, \|\sqrt{n\rho}\Psi^u\|^2$, 
so that $\Psi^{GA(\rho)}\sim\tilde\Lambda\, \Psi^{DA(\rho)}$, where $\tilde\Lambda$ is independent of $\Psi^{DA(\rho)}$ with $\PPP(\tilde\Lambda\in d\lambda)=\lambda^2\,\PPP(\Lambda\in d\lambda)$. When projecting to $\sphere(\Hilbert)$, the factor $\tilde\Lambda$ cancels out, so that $\Psi^{DAP(\rho)}\sim\Psi^{GAP(\rho)}$.

To see that the third definition is equivalent, note that $\Phi$ defines an (anti-linear) mapping $\Hilbert_2\to\Hilbert$ by $|\psi_2\rangle \mapsto \scp{\psi_2}{\Phi}$. To express this mapping explicitly in terms of the Schmidt decomposition of $\Phi$,
\be
\Phi = \sum_j \sqrt{p_j} |j\rangle |\phi_j\rangle
\ee
for some ONB $\{\phi_j:j=1\ldots n\}$ of $\Hilbert_2$, the vector $\psi_2=\sum_j c_j |\phi_j\rangle$ gets mapped to $\sum_j c_j^* \sqrt{p_j} |j\rangle$. That is, except for the conjugation, the mapping acts like $\sqrt{\rho}$. Thus, it maps the distribution $u_2$ to $D(\rho)$ and $\mu_2$ to $DA(\rho)$, except for a rescaling in $\Hilbert$ by a factor $\sqrt{n}$. The remaining step is the usual projection to $\sphere(\Hilbert)$, which also cancels the $\sqrt{n}$.

Here are further properties of $GAP$ measures. If $\rho$ is proportional to a projection, $\rho=(\dim W)^{-1}\, P_W$ for some subspace $W\subseteq \Hilbert$, then $GAP(\rho)=u_{\sphere(W)}$. In general, in a certain precise sense, $GAP(\rho)$ is the most spread-out distribution on $\sphere(\Hilbert)$ with density matrix $\rho$ \cite{JRW94}. Furthermore, $\rho\mapsto GAP(\rho)$ is covariant under unitary transformations $U$, i.e., $U\Psi^{GAP(\rho)}$ has distribution $GAP(U\rho U^{-1})$ \cite{gap2006}. The \emph{heredity} property  \cite[Section 3.5]{gap2006} says that if $\Psi\sim GAP(\rho_1\otimes\rho_2)$ then the conditional wave function $\psi_1^\cond$ has distribution $GAP(\rho_1)$. From the well-known fact of equivalence of ensembles, i.e., $\rho_{[E,E+\delta]} \approx \rho_{\beta}$ for macroscopic systems and suitable $\beta$, and the fact that $GAP(\rho)$ depends continuously on $\rho$, we obtain that
\be\label{uGAP}
u_{[E,E+\delta]}= GAP(\rho_{[E,E+\delta]}) \approx GAP(\rho_{\beta}) \, .
\ee
Applying this to the system $S\cup B$ considered in Section 1 with $\rho_\beta=\rho_\beta^{S\cup B}$, and neglecting the interaction between $S$ and $B$, it follows that if the wave function $\Psi$ of $S\cup B$ is random with distribution $u_{[E,E+\delta]}$, then the distribution $\mu^\cond_S$ of the conditional wave function $\psi^\cond_S$ is approximately $GAP(\rho^S_\beta)$. This statement is related to, but weaker than, Proposition~\ref{prop:GAP} above; the latter asserts that $\mu^\cond_S$ is actually near $GAP(\rho^S_\beta)$ for \emph{most} $\Psi$, not only \emph{on average}.

This completes our review of the $GAP$ measure.

\section{The Typical Distribution of the Conditional Density Matrix}

In this section, we derive and discuss the main result of our paper, Proposition~\ref{prop:condDM}, concerning the thermal equilibrium distribution of $\rho^{\cond}_S$.

\subsection{Derivation of Proposition~\ref{prop:condDM}}

Consider first $\psi^\cond_{S\cup s}$ and apply $GAP$ typicality, i.e., Proposition~\ref{prop:GAP}, to $S\cup s$ instead of $S$ and $y$ instead of $B$. Since, by assumption, the interaction between the $y$-system and $S\cup s$ is negligible, and since, by assumption, $\Hilbert_y$ and $\Hilbert_{[E,E+\delta]}$ have large dimensions, and $H_y$ has a reasonable distribution of eigenvalues, the hypotheses of Proposition~\ref{prop:GAP} are fulfilled. Thus, for most $\Psi$, $\psi^\cond_{S\cup s}\sim GAP(\rho_\beta^{S\cup s})$.

Now apply canonical typicality, i.e., Proposition~\ref{prop:cantyp}, to the $s$-system instead of $B$ and $\psi^\cond_{S\cup s}$ instead of $\Psi$; note that the operator called $\rho_S^\red=\tr_B |\Psi\rangle\langle\Psi|$ in Proposition~\ref{prop:cantyp} is then exactly $\rho_S^\cond$. By equivalence of ensembles as in \eqref{uGAP} and the large dimension of $\Hilbert_s$, what is true of most vectors in an energy shell of $\Hilbert_S\otimes\Hilbert_s$ is also true, with probability near 1, of a $GAP(\rho_\beta^{S\cup s})$-distributed vector in $\Hilbert_S\otimes\Hilbert_s$, such as $\psi^\cond_{S\cup s}$. Likewise, $\dim\Hilbert_{[E,E+\delta]}$ in Proposition~\ref{prop:cantyp} corresponds to the number of dimensions in $\Hilbert_S\otimes\Hilbert_s$ over which $GAP(\rho_\beta^{S\cup s})$ is spread out, and, again using equivalence of ensembles, the reasonable distribution of eigenvalues of $H_s$, and the large dimension of $\Hilbert_s$, this is large. By assumption, the interaction between $S$ and $s$ is negligible. Thus, the hypotheses of Proposition~\ref{prop:cantyp} are fulfilled, and we obtain that, for most $\Psi$ and with probability near 1, $\rho^\cond_S\approx \rho_\beta^S$. This is what we claimed.

We have a few remarks on this derivation. Since we did not aim at mathematical rigor, it is not surprising that the above derivation is not watertight. One loophole is that, in the known rigorous versions \cite{GLMTZ11} of Proposition~\ref{prop:GAP}, the required minimum dimensions of $\Hilbert_B$ and $\Hilbert_{[E,E+\delta]}$ depend on the dimension of $\Hilbert_S$; when we make both $y$ and $s$ larger, $\dim\Hilbert_{S\cup s}$ grows as well, and it is not obvious whether $\dim\Hilbert_y$ and $\dim\Hilbert_{[E,E+\delta]}$ will still be large enough compared to $\dim\Hilbert_{S\cup s}$ for the rigorous versions of Proposition~\ref{prop:GAP} to apply. However, we believe that Proposition~\ref{prop:GAP} is a robust statement that remains true in the regime in which we use it. In addition, a rigorous version of Proposition~\ref{prop:condDM} would require a careful examination of the equivalence of ensembles as we use it. 

As another remark on Proposition~\ref{prop:condDM}, we note that, given that $\nu^\cond_S$ is narrowly peaked, it is clear already before Proposition~\ref{prop:condDM} that the location of the peak must be $\rho_\beta$ because we knew from earlier that the average of $\rho^\cond_S$ is $\tr_{y\cup s}|\Psi\rangle\langle\Psi|$, which, by canonical typicality, is $\rho_\beta$.

\subsection{Alternative Derivation of Proposition~\ref{prop:condDM}}\label{sec:CLT}

Let us repeat Eq.~\eqref{rhocondrx}:
\be\label{rhocondrx2}
\rho^{\cond}_{S\:\:\:r,r'} (x,x') = \frac{1}{\mathcal{N}(Y)}\sum_{s} \Psi_{r,s}(x,Y)\, \Psi^*_{r',s}(x',Y) \; .
\ee
It is also possible to arrive at Proposition~\ref{prop:condDM} by applying the central limit theorem (or at least its spirit) to this (or an equivalent) expression, which provides $\rho^\cond_S$ as a sum $\sum_s$ over many (since $\dim\Hilbert_s$ is large) random terms. We will explain how these terms can, under certain conditions, be written as independent random variables. The central limit theorem (e.g., \cite[Thm.~27.4]{Bill}) then tells us that the matrix entries of $\rho^\cond_S$ have a Gaussian distribution; the most relevant conclusion for us, however, does not even require the central limit theorem but is based simply on computing the variance of the latter distribution, which, we will show, is very small compared to (the square of) its mean. Thus, we will conclude, the distribution is narrowly peaked, and we obtain an estimate of how narrow.

Here is the derivation. We want to show that for most $\Psi$ and most ONBs $\{|y\rangle\}$, the probability is near 1 that
\be\label{rhocondrhobeta}
\rho^\cond_S\approx \rho_\beta^S\,.
\ee
Equivalently, we can regard $\Psi$ as random with distribution $u_{[E,E+\delta]}$, $\{|y\rangle\}$ as random and independent of $\Psi$ with uniform distribution, and claim that the probability of \eqref{rhocondrhobeta} is near 1. Since the ONB $\{|y\rangle\}$ is uniformly distributed, the distribution $\|\scp{y}{\Psi}\|^2$ is more or less uniform over the $y$s, and thus the claim that (for most $\Psi$ and most ONBs) most $y$ \emph{with respect to $\|\scp{y}{\Psi}\|^2$} have a property $p$ is equivalent to the claim that (for most $\Psi$ and most ONBs) most $y$ \emph{with respect to the uniform distribution} have property $p$. So, from now on we consider the uniform distribution for $y$. We will show that for most $\Psi$, most ONBs, and most $y$ (or, equivalently,\footnote{The point here is that we can change the order of the quantifiers (``for most $y$'' etc.) without changing the content of the statement; this was not possible as long as the notion of ``most $y$'' depended on $\Psi$.} for most $y$, most ONBs, and most $\Psi$), \eqref{rhocondrhobeta} holds. In fact, we derive the (seemingly stronger but actually equivalent) statement that for every $y$, most ONBs, and most $\Psi$, \eqref{rhocondrhobeta} holds. The latter is equivalent to the claim that for most unit vectors $|Y\rangle$ in $\Hilbert_y$ and random $\Psi$, the probability of \eqref{rhocondrhobeta} is near 1. That is what we will show.

The expression \eqref{rhoconddef2}, which we repeat here for convenience,
\be\label{rhoconddef3}
\rho^\cond_S=\frac{\tr_s \scp{Y}{\Psi} \scp{\Psi}{Y}}{\bigl\| \scp{Y}{\Psi}\bigr\|^2}\,,
\ee
remains unchanged if we change $\Psi$ by a positive factor $\Lambda$ that is random but independent of $\Psi$. In particular, we can replace the $u_{[E,E+\delta]}$-distributed $\Psi$ by a $G(\rho_{[E,E+\delta]})$-distributed vector $\tilde\Phi$. By equivalence of ensembles, since $B$ is a macroscopic system, $\rho_{[E,E+\delta]}\approx \rho^{S\cup B}_\beta$, and since interaction is negligible, $\rho_\beta^{S\cup B}\approx \rho_\beta^S\otimes \rho_\beta^s\otimes \rho_\beta^y$; so we assume
\be\label{tildePhiGrhobeta}
\tilde\Phi\sim G\Bigl(\rho_\beta^S\otimes \rho_\beta^s\otimes \rho_\beta^y\Bigr)\,.
\ee
It follows that, for any fixed vector $|Y\rangle\in\sphere(\Hilbert_y)$, $\Phi:= \scp{Y}{\tilde\Phi}$ (where $\scp{\cdot}{\cdot}$ is a partial inner product) has distribution $G\bigl(\rho_\beta^S\otimes\rho_\beta^s \scp{Y}{\rho_\beta^y|Y}\bigr)$. Since a fixed positive stretching factor such as $\scp{Y}{\rho_\beta^y|Y}$ does not affect $\rho^\cond_S$, we can assume
\be\label{PhiG}
\Phi\sim G\bigl(\rho_\beta^S\otimes\rho_\beta^s\bigr)
\ee
and obtain
\be\label{rhotrPhi}
\rho^\cond_S \approx \|\Phi\|^{-2}\, \tr_s |\Phi\rangle\langle\Phi| \,.
\ee
In fact $\|\Phi\|^2\approx 1$ with probability near 1 because $\EEE\|\Psi^{G(C)}\|^2=\tr C$ and, by the law of large numbers using that $\Hilbert_S\otimes\Hilbert_s$ has high dimension, the value of $\|\Psi^{G(C)}\|^2$ is close to its expectation with probability near 1.\footnote{\label{fn:Es}Actually, we use here more than just large $\dim\Hilbert_s$; we use that among the eigenvalues of $\rho_\beta^S\otimes\rho_\beta^s$ are not just a few dominating ones while all others are negligible, but that many of them are of comparable size; that is the case since $H_s$ has a reasonable distribution of eigenvalues.} So we can drop the factor $\|\Phi\|^{-2}$ in \eqref{rhotrPhi}.

Since the trace can be expressed in any ONB of $\Hilbert_s$, it is convenient to choose an eigenbasis of $\rho_\beta^s$ (i.e., an energy eigenbasis), which we denote $|s\rangle$, so
\be
\rho^\cond_S \approx \sum_s \scp{s}{\Phi} \scp{\Phi}{s}\,.
\ee

It suffices for proving \eqref{rhocondrhobeta} to show that
\be\label{rhocondrhobeta2}
\scp{\phi}{\rho^\cond_S|\phi} \approx \scp{\phi}{\rho_\beta^S|\phi}
\ee
for all $\phi\in\sphere(\Hilbert_S)$. So fix a $\phi$. We need to evaluate
\be\label{phirhos}
\scp{\phi}{\rho^\cond_S|\phi} \approx \sum_s \scp{\phi\otimes s}{\Phi} \scp{\Phi}{\phi\otimes s}\,.
\ee
Note that in this expression, the inner products are complex-valued (as opposed to partial inner products). By \eqref{PhiG}, the $X_s:= \scp{\phi\otimes s}{\Phi}$ are jointly Gaussian complex random variables with covariance matrix $\scp{\phi}{\rho_\beta^S|\phi}\, \rho_\beta^s$; since $\{|s\rangle\}$ is an eigenbasis of this operator, the $X_s$ are independent, and each $X_s$ has mean 0 and variance
\be
\sigma_s^2=\scp{\phi}{\rho_\beta^S|\phi}\,Z(\beta)^{-1}\exp(-\beta E_s)\,,
\ee
where $E_s$ is the energy eigenvalue, $H|s\rangle=E_s|s\rangle$, $H$ is the Hamiltonian of the $s$-system, and $Z(\beta)= \tr\exp(-\beta H)$ is its partition function.

For any complex Gaussian random variable $X$ with mean 0 and variance $\sigma^2$, we have that
\be
\EEE|X|^2 = \sigma^2\,, \quad
\Var(|X|^2)=\sigma^4\,.
\ee
(Indeed, $|X|^2$ is exponentially distributed with expectation $\sigma^2$, and it is well known that, for exponentially distributed $Y$, $\Var(Y)=(\EEE Y)^2$.)

Since $\EEE$ is additive and, for independent variables, so is $\Var$, we obtain from \eqref{phirhos} that
\be\label{E}
\EEE \scp{\phi}{\rho^\cond_S|\phi} 
\approx \sum_s \scp{\phi}{\rho_\beta^S|\phi}\,Z(\beta)^{-1}\, \exp(-\beta E_s) 
=  \scp{\phi}{\rho_\beta^S|\phi}
\ee
and
\be\label{Var}
\Var \scp{\phi}{\rho^\cond_S|\phi} 
\approx \sum_s \scp{\phi}{\rho_\beta^S|\phi}^2 \,Z(\beta)^{-2}\, \exp(-2\beta E_s) 
= \scp{\phi}{\rho_\beta^S|\phi}^2 \frac{Z(2\beta)}{Z(\beta)^2}\,.
\ee
The central limit theorem now tells us that the positive random quantity $\scp{\phi}{\rho^\cond_S|\phi}$ has an approximately Gaussian distribution with expectation \eqref{E} and variance \eqref{Var}. The relevant point for us is that this distribution is very narrow, i.e., that the variance is small compared to the square of the expectation. Indeed,
\be\label{VarE2}
\frac{\Var}{\EEE^2} \approx \frac{Z(2\beta)}{Z(\beta)^2}=\tr \bigl[(\rho_\beta^s)^2 \bigr]\,,
\ee
which is very small provided that among the eigenvalues of $\rho_\beta^s$ are not just a few dominating ones while all others are negligible, but that many of them are of comparable size. As mentioned already in Footnote~\ref{fn:Es}, that is the case since $H_s$ has a reasonable distribution of eigenvalues. The quantity \eqref{VarE2} is of the order $1/n$ with $n$ the number of eigenvalues of $\rho_\beta^s$ of the size that contributes most to $Z(\beta)$; we expect that $1/n$ should be of order of magnitude $\exp(-N)$ with $N$ the number of particles in the heat bath. Anyway, if all eigenvalues are equal then \eqref{VarE2} yields $(\dim\Hilbert_s)^{-1}$, which is also of the order of magnitude $\exp(-N)$. This completes the second derivation of Proposition~\ref{prop:condDM}.

To sum up, suppose that $S\cup B$ is in thermal equilibrium, that $y$ and $s$ in $B=y \cup s$ each have many degrees of freedom, and that
interaction is small. Then the conditional density matrix of $S$ is canonical.

\bigskip

\noindent\textit{Acknowledgments.} 
R.T.\ was supported by grant no.\ 37433 from the John Templeton Foundation. 

\bibliographystyle{plain}
\bibliography{references}

\end{document}